\documentclass[prl,aps,twocolumn,showpacs,nobibnotes,epsf]{revtex4}

\usepackage{graphicx}
\usepackage{dcolumn}
\usepackage{bm}
\usepackage{SIunits}
\usepackage{verbatim}
\usepackage{placeins}

\begin{document}
\title{Large magnetic anisotropy in $Fe_xTaS_2$ single crystals}
\author{G. Wu$^1$$^{\dag}$, B. L. Kang $^1$, Y.-L. Li$^{2,3}$, T. Wu$^{1,5}$, N. Z. Wang$^1$, X. G. Luo$^1$, Z. Sun$^{4,5}$, L.-J. Zou$^2$, X. H. Chen$^{1,5}$}
\altaffiliation{Corresponding author} \email{chenxh@ustc.edu.cn}
\altaffiliation{$^{\dag}$ Present address: Oxford
Instruments(Shanghai) Co., Ltd, Shanghai 200233, China}
\affiliation{ $^1$Hefei National Laboratory for Physical Science at
Microscale and Department of Physics, and Key Laboratory of
Strongly-coupled Quantum Matter Physics, Chinese Academy of
Sciences,
University of Science and Technology of China, Hefei, Anhui 230026, China \\
 $^2$ Key Laboratory of Materials Physics, Institute of Solid State
 Physics, Chinese Academy of Sciences, P. O. Box 1129, Hefei, Anhui
 230031, China\\$^3$ Laboratory for Quantum Design of Functional Materials, Jiangsu
Normal University, Xuzhou, Jiangsu 221116, China\\
$^4$National Synchrotron Radiation Laboratory, University of
Science and Technology of China, Hefei, Anhui 230026, China\\
 $^5$ Collaborative Innovation Center of Advanced Microstructures, Nanjing University, Nanjing 210093, China}

\begin{abstract}
In intercalated transition metal dichalcogenide $Fe_xTaS_2$ (0.2
$\leq$ x $\leq$ 0.4) single crystals, large magnetic anisotropy is
observed. Transport property measurements indicate that heavy
Fe-doping leads to a large anisotropy of resistivity
($\rho$$_{c}$/$\rho$$_{ab}$). A sharp M-H hysteresis curve is
observed with magnetic field along c-axis, while a linear
magnetization appears with magnetic field applied in the ab-plane.
The angular dependent magnetic susceptibility from in-plane to
out-of-plane indicates that magnetic moments are strongly pinned
along the c-axis in an unconventional manner and the coercive field
reaches as large as 6 T at T = 5 K. First-principles calculation
clearly suggests that the strong spin-orbital coupling give rise to
such a large anisotropy of magnetism. The strong pinning effect of
magnetic moments along c-axis makes this material a very promising
candidate for the development of spin-aligner in spintronics
devices.
\end{abstract}


\vskip 300 pt

\maketitle

Transition metal dichalcogenides have been subject to intensive
study due to their rich physical properties as a result of strong
electron correlation and electron-phonon coupling.
These intriguing compounds share the same
characteristics that they are all  low dimensional with layered structures, and
receptive to intercalation due to van der Waals force between S-M-S
layers\cite{Whittingham, Whittingham2, Trichet}. In this family,
people have discovered charge density wave (CDW) order in
1T-$TaS_2$\cite{ref1}, long range magnetic order in $Mn_xTaS_2$
(ferromagnetism)\cite{ref2} and $Fe_xTiSe_2$
(antiferromagnetism)\cite{ref3}, and superconductivity in
$K_xTiSe_2$ and $Cu_xTiSe_2$\cite{ref4, Morosan, Wu}.
Among them, iron-intercalated
$1T-TaS_2$ compounds have been greatly attractive
owing to their exotic
magnetic properties originating from the interaction between crystal
field, electron orbit and local moment of $Fe^{2+}$.

Previous works by Cava {\it et al.}\cite{cava} and Ong {\it et
al.}\cite{ong} have shown that in $Fe_xTaS_2$ (x = 1/4) $Fe^{2+}$
ions are distributed periodically in a doubled cell of parent
compound $TaS_2$ rather than randomly scattered between $TaS_2$
layers. Computational results has validated the ferromagnetic
transition for x = 0.33\cite{Dijkstra}. In Fe$_x$TaS$_2$ system, the
magnetic properties of Fe$_x$TaS$_2$ change from spin glass
(x$<$0.2) to ferromagnetism (0.2$\leq$x$\leq$0.4)\cite{Eibschutz}
and antiferromagnetism(x$>$0.4)\cite{Narita} with the increasing
of x. For 0.20$\leq$x$\leq$0.40, the Curie temperature T$_c$ is
non-monotonous with x,  T$_c$=90 K for x=0.2; T$_c$=163 K
for x=0.26;  T$_c$=55 K for x=0.34\cite{Eibschutz}. The Curie
temperature decreases as x further increases. Cheong {\it et al.}
suggested the nano domains and unqueched orbital moment contribute to the large magnetic anisotropy and magnetoresistance, respectively \cite{Choi,Ko}. However, Morosan {\it et al.} attributed the large
magnetoresistance to disorder \cite{Chen}. A systematical study on
the magnetic and transport properties of $Fe_xTaS_2$, especially the
magnetization and transport behaviors under different magnetic
fields, are critical to reveal the intrinsic properties of this material. In this paper, we presence
the large magnetic anisotropy in the resistivity, magnetic
susceptibility and magnetization in Fe$_x$TaS$_2$ single crystals
and elucidate the microscopic origin of such a huge magnetic
anisotropy.

The Fe$_{x}$TaS$_{2}$ (0.2$\leq$x$\leq$0.40) single crystals were
grown by the chemical iodine vapor transport method. Fe ($>$99.5\%),
Ta ($>$99.5\%) and S ($>$99.5\%) element powders were mixed and
thoroughly ground, then pressed into pellets. The pellets were
sealed under vacuum ($<$1.0$\times$10$^{-2}$Pa) in a quartz tube
($\phi$ 13mm$\times$150mm) with a small quantity of iodine
(10mg/cm$^{-3}$). The tube was slowly heated to 1000$^o$C in 400
minutes, and the cold end of the tube was kept at 950$^o$C. After
ten days, the furnace was cooled to room temperature over a few
hours. Finally many high quality plate-like Fe$_{x}$TaS$_{2}$ single
crystals were obtained. The single crystals were black and the
typical dimension is about 3$\times$3$\times$0.3mm$^3$.

\begin{figure}[htbp]
\centering
\includegraphics[width=0.5\textwidth]{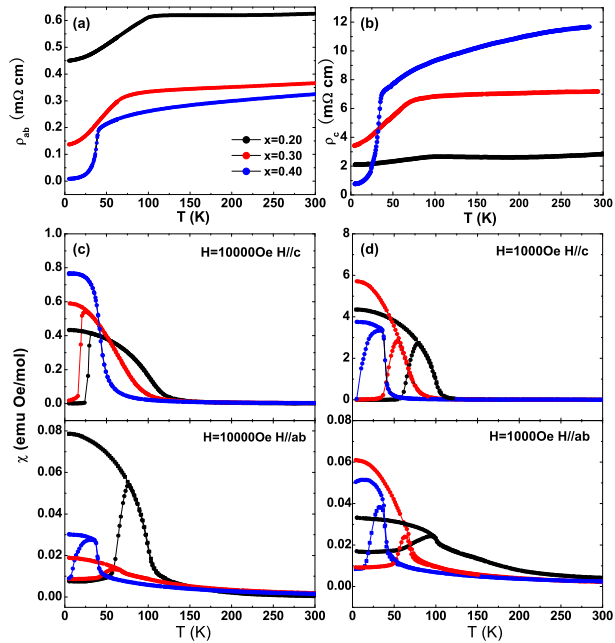}
\caption{(Color online) (a): Temperature dependence of in-plane
resistivity; (b): Temperature dependence of out-of-plane
resistivity; (c): Temperature dependence of magnetic susceptibility
under the magnetic field of 1 Tesla applied along in-plane and
out-of plane, respectively; (d):Temperature dependence of magnetic
susceptibility under the magnetic field of 0.1 Tesla applied along
in-plane and out-of plane, respectively. The data were
obtained from the $Fe_xTaS_2$ single crystals with x = 0.2 (black),
0.3 (red) and 0.4 (blue). }\label{fig1}
\end{figure}

Figs. 1(a) and (b) demonstrates how the resistivity anisotropy
evolves with Fe doping. As the dopant Fe increases from 0.2 to 0.4,
the electronic resistivity in the ab-plane significantly decreases,
while the out-of plane resistivity increases drastically, thus leading to
an enhanced anisotropy of $\rho_c/\rho_{ab}$. The ferromagnetic phase
transition indicated by the kinks in resistivity is gradually suppressed with Fe doping,
from 105 K for x=0.2, to 76 K for x=0.3, and 43
K for x=0.4, respectively. Figs. 1(c) and (d) show the temperature
dependence of magnetic susceptibility with the applied magnetic
field of 10000 Oe and 1000 Oe, respectively. The in-plane
susceptibility is significantly smaller than the c-axis
susceptibility, unambiguously indicating that the easy axis is
perpendicular to the ab-plane, and a large field tends to reduce such
magnetic anisotropy. It should be pointed out that the zero field cooled
magnetization curve with the magnetic field applied along c-axis
reaches almost zero below $T_c$, while magnetization with the
magnetic field applied in the ab-plane can not yields zero
susceptibility below $T_c$. This contrast implies that there exists an
anisotropic magnetic energy between the ab-plane and c-axis. These
results are consistent with those reported in Ref. \cite{Eibschutz}

\begin{figure*}[htbp]
\centering
\includegraphics[width=0.9\textwidth]{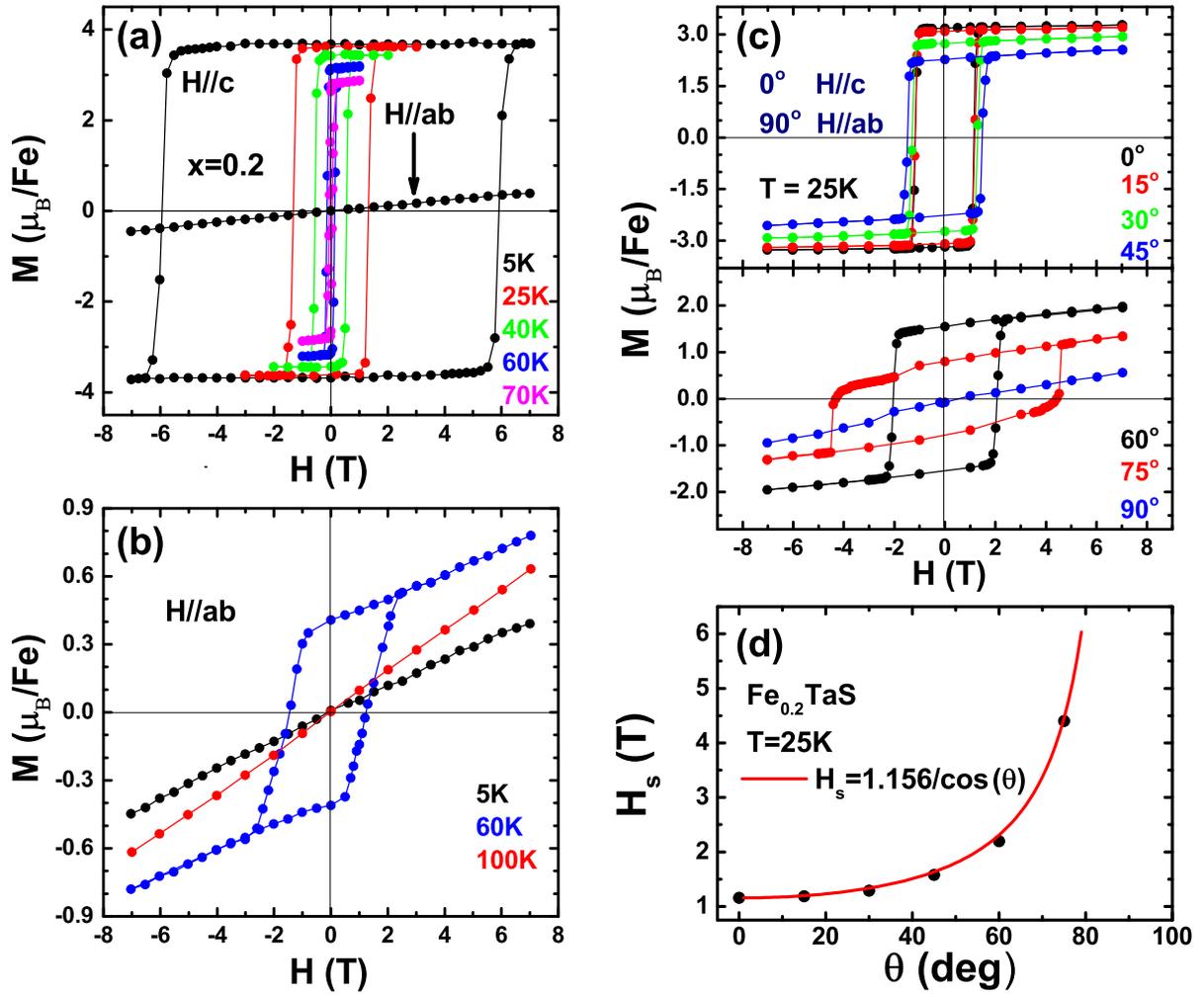}
\caption{(Color online) Temperature dependence of M-H loop in
$Fe_{0.2}TaS_2$ with the external magnetic field applied (a)
along c-axis and (b) in ab-plane. (c):  M-H loops at 25 K in
$Fe_{0.2}TaS_2$ with the external magnetic field applied along
different directions from c-axis to ab-plane. (d): The
angle-dependent the switching field $H_s$ of magnetization. The red
curve is the fitting data obtained  by the formula
$H_s$=1.156/cos($\theta$).} \label{fig2}
\end{figure*}

Fig. 2(a) shows the magnetization loops at different temperatures
under the magnetic field applied along c-axis of Fe$_{0.2}$TaS$_{2}$. Clear M-H loops shows up below the
Curie temperature with the external magnetic field applied along
the c-axis. The coercive field systematically
decreases with increasing the temperature. The most striking
features are the large coercive field up to 6 T at 5 K and the
sharp magnetic transition to reach saturation magnetization ($M_s$).
Fig. 2(b) shows the evolution of the magnetization ($M_{ab}$) with
the external magnetic field applied in ab-plane at different
temperatures. There is no M-H loop at 5 K, and the magnetization
$M_{ab}$ increases linearly with the magnetic field up to 7 T with a small
slope. This behavior suggests that the pinning force along c-axis is so strong
at 5 K that only statistical net moment could be observed in the
ab-plane magnetization ($M_{ab}$).
When temperature rises to 60 K, the magnetic anisotropy decreases
and more local moments from Fe ions are aligned in the ab-plane to
form domain, so that a skew M-H loop appears during the
magnetizing process as shown in Fig. 2(b). Such a behavior implies that the
in-plane magnetic crystal energy is weaker than that along
c-axis. As temperature rises to above $T_{c}$, the thermal fluctuation becomes dominant, and the
magnetization correspondingly follows a linear dependence of the
magnetic field and shows a paramagnetic behavior. As shown in Fig. 2(a), the slope of the linear magnetic
field dependent $M_{ab}$ is so small
that the anisotropy field $H_A$ approximately exceeds 60 T if we extrapolate the linear $M_{ab}$ to the saturation value of
$M_{c}$. Therefore, the first order anisotropic coefficient
$K_1=\mu_0M_sH_A$/2 is estimated to be about 6.5 meV. As temperature
increases, the coercive field and remaining
magnetization shrink quickly, suggesting a rapid decrease of $K_1$.

Fig. 2(c) shows the evolution of
magnetization loop with the direction of external magnetic field at T = 25 K.
A continuous 'flattened' loop is observed, and the saturated
magnetization continuously decreases and the coercive field
monotonously increases with the variation of the applied
magnetic field from along the c-axis to the in-plane.
The switching
field $H_s$ of the magnetization obtained from Fig. 2(c) are plotted
in Fig. 2(d), and the evolution of $H_s$ with the
direction of the applied magnetic field  can be well fitted by the
formula $H_s$=1.156/cos($\theta$). This behavior indicates that
$Fe_{0.2}TaS_{2}$ is almost an ideal single domain ferromagnetic
system with the easy axis perpendicular to the ab-plane.

\begin{figure}[htbp]
\centering
\includegraphics[width=0.45\textwidth]{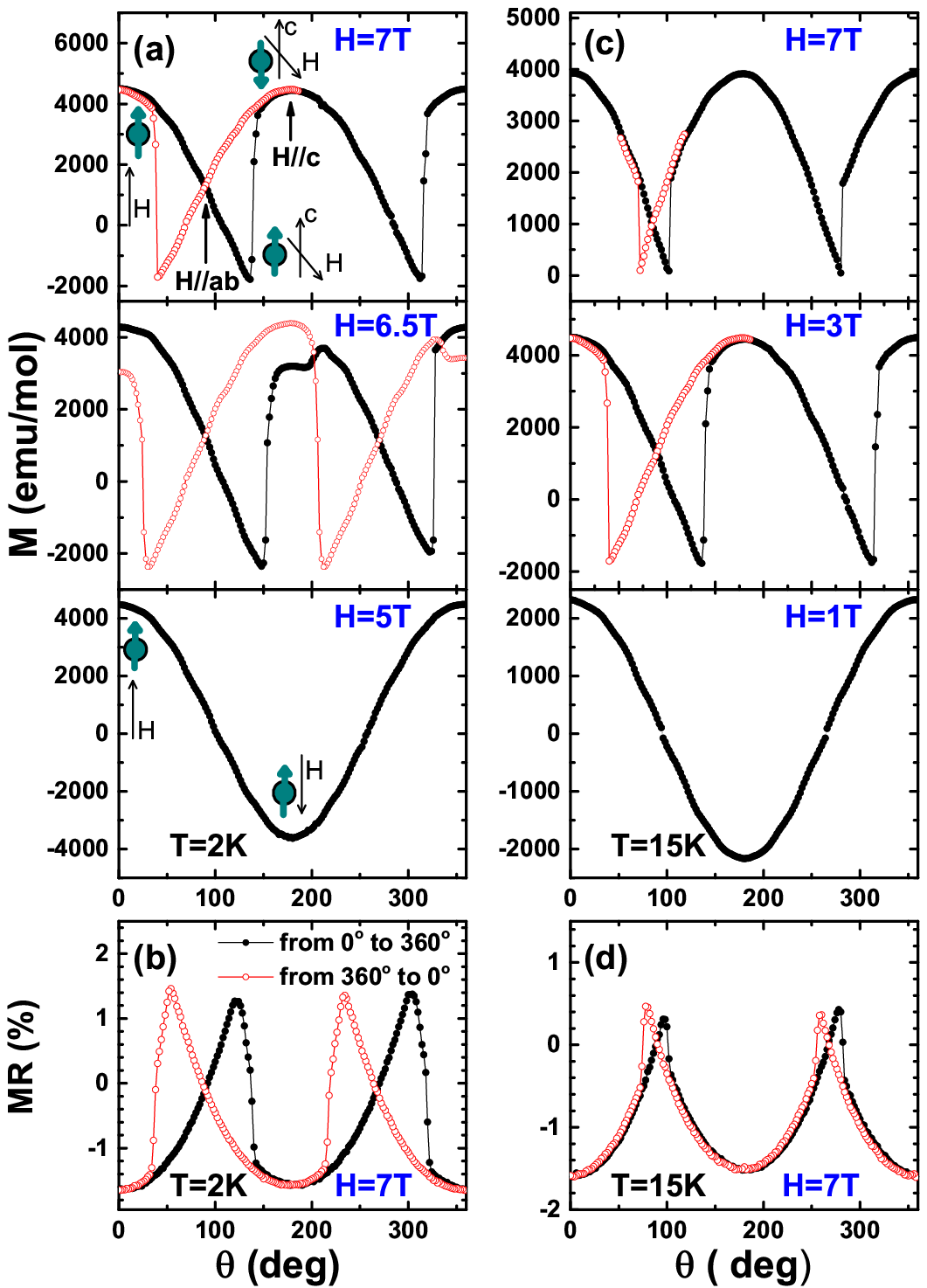}
\caption{(Color online) Angle-dependent magnetization under
different magnetic fields by tuning the external magnetic field from
along c-axis to in the ab-plane for $Fe_{0.2}TaS_2$ single crystal
at (a): 2 K (c): 15 K. The green dot and arrow represent the
direction of totally local moment of Fe ions. Angle-dependent
magnetoresistance under the magnetic field of 7 Tesla by tuning the
magnetic field from along c-axis to in the ab-plane for
$Fe_{0.2}TaS_2$ single crystal at (a): 2 K (c): 15 K. The angle of
 $0^o$ represents the direction of magnetic field along c-axis direction, while the
 angle of $90^o$ means the magnetic field in ab-plane. The black data are obtained by
 tuning the H from $0^o$ to $360^o$£¬ while the red data are obtained  by
 tuning the H from $360^o$ to $0^o$. The black and red data show a perfect symmetry relative to ab-plane. }\label{fig3}
\end{figure}

In order to quantify the pinning strength in this system,
angle-dependent magnetization and magnetoresistance with the direction of
external magnetic field varying from c-axis to ab-plane  were carried out, and
the results are shown in Fig. 3. As shown in Fig. 3(a), the applied external field up to 5 Tesla still cannot flip the
spins at T = 2 K, so that the magnetization behaves exactly in a sinusoid way.
As the external magnetic field rises to 6.5 and 7 Tesla, there
are clear jumps on magnetization at $155^o$ and $140^o$ relative
to c-axis, respectively. In Fig. 3(c),
similar behavior persists at T = 15 K.
The magnetic field up to 1 T can not flip the spins, while the spin
flip takes place with increasing the field up to 3 Tesla. Our data indicates that
the \emph{critical field} of the spin flip decreases with increasing
temperature. A pictorial explanation is sketched in
Fig. 3(a). The spins of Fe ions are pinned along c-axis at low
temperature. When the external magnetic field is rotated, the spin
alignment holds until the magnetic energy $\mu
H$cos($\theta$) surpasses the pinning energy $E_p$, then an
180-degree flip happens for all spins, leading to the sharp jump of
magnetization from negative (positive) to positive (negative). Based
on the critical angle of the spin flip at certain magnetic field, we
can estimate the pinning energy in terms of following relation:
\begin{equation}
E_p = -\mu Hcos(\theta)
\end{equation}
which could be verified for both of the cases at 2 K and 15 K,
respectively. The pining energy $E_p$ is $\sim$ 1.24 meV at 2 K, and
 $\sim$ 0.53 meV at 15 K. $E_p$ gradually decreases with increasing temperature
due to thermal fluctuation.  Similar magnetoresistance measurement
confirms the spin flip under rotating external field from the c-axis to  the
ab-plane, as shown in Figs. 3(b) and 3(d). The flipping
critical angle is exactly the same as those obtained from the
angle-dependent magnetization. These results not only confirm the existence
of pinning along the c-axis, but also suggest that magnetic
scattering is profound in this system. Moreover, as temperature
rises, the sharp jump in magnetization become week and finally
disappears, closely correlated with the behavior of magnetoresistance. Such a correlation indicates that the magnetic
scattering is an indispensable component in the electronic transport
and there exsits a strong spin-electron coupling.

\begin{figure}[htbp]
\centering
\includegraphics[width=0.45\textwidth]{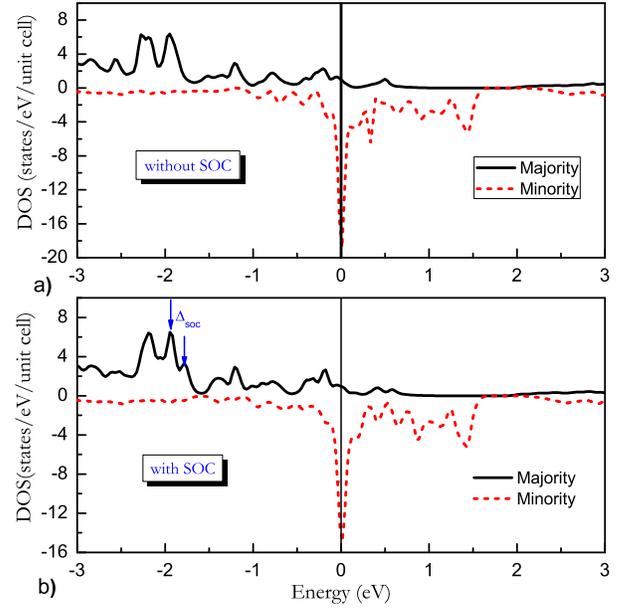}
\caption{(Color online) The spin-polarized density of states
(DOS) in $Fe_{0.25}TaS_2$, calculated without (a) and with (b) spin-orbit coupling. The DOS near the Fermi
energy is highly spin-polarized, with dominant $d_{x^2-y^2}$ electrons. The
arrows in (b) indicate the energy splitting caused by the spin-orbit
coupling.}\label{fig4}
\end{figure}

The key to the understanding of these exotic phenomena in $Fe_xTaS_2$ system
is to clarify the roles of Fe spins and its coupling with
environment. In conventional 3d transition metals, the orbital
angular momentum is partially or almost fully quenched due to the
complete or partial lift of the orbital degeneracy in 3d electronic
states, as a result of the presence of low symmetric crystal field.
In the present case, our theoretical calculations and analysis show
that the intercalated Fe ions are coordinated in a triangular
crystalline field of six S ions, the orbital degeneracy of Fe 3d
orbitals is partially removed. A crystal field separates the upper
two-fold e$_g$ levels from lower three-fold degenerate t$_{2g}$ levels
with $\Delta_{CF}$ of 1 eV. This leads to an effective orbital
angular momentum L = 1, in agreement with Cheong {\it et al}.'s
result \cite{Ko}. Considering the Fe spin of S=2 in the Fe 3d$^4$
configuration, we have the total angular momentum of J = 3 for Fe
ions. Therefore, the theoretical value of magnetic moment in
$Fe^{2+}$ is about 5 $\mu_B$, in good agreement with our and Cava
{\it et al.}'s \cite{cava} experimental results. In order to find
out the origin of strong pinning of spin alignment along c-axis, we
have carried out further first-principles electronic structure calculation,
and found considerable effect from the spin-orbital coupling, as
shown in Fig. 4. By comparing the density of states (DOS) without and
with the spin-orbital coupling, one may find that the spin-orbital
coupling brings about a 0.05 eV splitting in the $d_{x^2-y^2}$
orbital around 2 eV below the Fermi level ( see Figs. 4(a) and
4(b)). The strongly spin-polarized Fermi surface also directly gives
rise to the strong field-dependent transport properties. At last, we could
infer the magnetic anisotropy parameter to be
\begin{equation}
K_1 = \frac{25\lambda^2}{9\Delta_{CF}} \simeq 7 meV
\end{equation}
which decently agrees with the experimental data of $\sim$ 6.5 meV.
We also notice that a recent work suggested the crucial role of
electronic correlation \cite{Zhu}.
The first-principles electronic structure calculation confirms that
the spin-orbital coupling can induced large magnetic anisotropy and
strong spin pinning along c-axis in $Fe_xTaS_2$, the calculated
magnetic anisotropy energy $K_1$ is well consistent with the
experiment result.

Efficient electrical injection of spin-polarized carriers from a
contact into a semiconductor is one of the essential requirements to
utilize carrier spin as an operational paradigm for future
electronic devices like a spin-LED\cite{Fiederling, Hanbicki, Liu}.
Two kinds of spin aligner are used to polarize the spin of the
carriers. One is diluted magnetic semiconductor and the other is
ferromagnetic metal. The large coercive field and the sharp
transition in the c-axis magnetization of $Fe_{x}TaS_2$ can provide
as the spin aligner in electrical spin injection in spintronics.

In summary, we have observed large magnetic anisotropy in
$Fe_xTaS_2$ system. Coercive field as large as 6 T and sharp
transition exist in c-axis magnetization. Angle-dependent
magnetization at different temperatures from ab-plane to c-axis
reveals that the strong pinning along c-axis fits the simple spin-flip model, in favor of the single
domain picture\cite{cava}. Theoretical analysis suggests that the magnetic anisotropy comes from the spin-orbital coupling of Fe ions
in the triangular crystal field. Both of first-principles
electronic structure calculation and experiment indicate a magnetic anisotropy
energy of $\sim$7 meV. Our data suggests that $Fe_xTaS_2$ can serve as
a spin aligner in electrical spin injection in
spintronics\cite{Fiederling}. The spontaneous strong spin
polarization in this material can also be adopted as a reliable spin
current source.

\textbf{Acknowledgements}
This work is supported by the National Key R$\&$D Program of the MOST of China (Grant No. 2016YFA0300201), the Nature Science Foundation of China (Grant No:
11474287), Hefei Science Center CAS (2016HSC-IU001) and the ``Strategic Priority Research Program'' of the Chinese Academy of Sciences (grant no. XDB04040100).

\end{document}